\documentclass[aps,prl,preprint,longbibliography,floats,floatfix,onecolumn,superscriptaddress]{revtex4-1}

\usepackage{latexsym,amsmath}
\usepackage{amsbsy}
\usepackage[pdftex]{graphicx}
\usepackage{bm}
\usepackage{amssymb}

\begin{document}

\title{Essential plasticity and redundancy of metabolism unveiled\\by synthetic lethality analysis}

\author{Oriol G\"uell}
\email{oguell@ub.edu}
\affiliation{Departament de Qu\'imica F\'isica, Universitat de Barcelona, Mart\'i i Franqu\`es 1, 08028, Barcelona, Spain}
\author{Francesc Sagu\'es}
\affiliation{Departament de Qu\'imica F\'isica, Universitat de Barcelona, Mart\'i i Franqu\`es 1, 08028, Barcelona, Spain}
\author{M. \'Angeles Serrano}
\affiliation{Departament de F{\'\i}sica Fonamental, Universitat de Barcelona, Mart\'{\i} i Franqu\`es 1, 08028 Barcelona, Spain}

\begin{abstract}
We unravel how functional plasticity and redundancy are essential mechanisms underlying the ability to survive of metabolic networks. We perform an exhaustive computational screening of synthetic lethal reaction pairs in {\it Escherichia coli} in a minimal medium and we find that synthetic lethal pairs divide in two different groups depending on whether the synthetic lethal interaction works as a backup or as a parallel use mechanism, the first corresponding to essential plasticity and the second to essential redundancy. In {\it E. coli}, the analysis of pathways entanglement through essential redundancy supports the view that synthetic lethality affects preferentially a single function or pathway. In contrast, essential plasticity, the dominant class, tends to be inter-pathway but strongly localized and unveils Cell Envelope Biosynthesis as an essential backup for Membrane Lipid Metabolism. When comparing {\it E. coli} and {\it Mycoplasma pneumoniae}, we find that the metabolic networks of the two organisms exhibit a large difference in the relative importance of plasticity and redundancy which is consistent with the conjecture that plasticity is a sophisticated mechanism that requires a complex organization. Finally, coessential reaction pairs are explored in different environmental conditions to uncover the interplay between the two mechanisms. We find that synthetic lethal interactions and their classification in plasticity and redundancy are basically insensitive to medium composition, and are highly conserved even when the environment is enriched with nonessential compounds or overconstrained to decrease maximum biomass formation.
\end{abstract}

\maketitle
\section{Introduction}
Homeostasis in living systems balances internal states over a possibly wide range of internal or external variations, and organisms unable to maintain stability in front of internal disruptions or changing environments can experience dysfunction and even collapse. At the level of cellular metabolism, multiple regulatory mechanisms control homeostasis, including allosteric or posttranslational activation of enzymes and modulation of enzymatic activity by transcriptional regulation~\cite{Desvergne:2006}. Research efforts have typically focused on elucidating the molecular basis of such controls, but we are still far from understanding the complex functional strategies that explain homeostasis at a systems level, even when this concept is loosened to that of maintaining viability in front of perturbations.  

Beyond the molecular level, plasticity and redundancy are large-scale strategies that offer the organism the ability to exhibit no or only mild phenotypic variation in front of environmental changes or upon malfunction of some of its parts. In particular, these mechanisms protect metabolism against the effects of single enzyme-coding gene mutations or reaction failures such that most metabolic genes are not essential for cell viability. However, some mutants fail when an additional gene is knocked out, so that specific pair combinations of non-essential metabolic genes or reactions become essential for biomass formation. As an example, double mutants defective in the two different phosphoribosylglycinamide transformylases present in {\it Escherichia coli} --with catalytic action in purine biosynthesis and thus important as crucial components of DNA, RNA or ATP-- require exogenously added purine for growth, while single knockout mutants do not result in purine auxotrophy~\cite{Smith:1993}.

These synthetic lethal (SL) combinations~\cite{Hartman:2001,Tucker:2003,Masel:2009,Nijman:2011} have recently attracted attention because of their utility for identifying functional associations between gene functions and, in the context of human genome, for the prospects of new targets in drug development. However, inviable synthetic lethal mutants are difficult to characterize experimentally despite the high-throughput techniques developed recently~\cite{Kaelin:2005}. We are still far from a comprehensive empirical identification of all SL metabolic gene or reaction pairs in a particular organism~\cite{Suthers:2009}, even more when considering different growth conditions. Metabolic screening based on computational methods, such as Flux Balance Analysis (FBA) simulating medium dependent optimal growth phenotypes, becomes then a powerful complementary technique particularly suited for an exhaustive {\it in silico} prediction of SL pairs~\cite{Suthers:2009,Deutscher:2006} in high-quality genome-scale metabolic reconstructions~\cite{Palsson:2006}.

With this perspective, we unveil how functional plasticity and redundancy are essential systems-level mechanisms underlying the viability of metabolic networks. In previous works on cellular metabolism~\cite{Almaas:2005,Harrison:2007}, plasticity was some times associated to changes in the fluxes of reactions when an organism is shifted from one growth condition to another. Instead, we discuss here functional plasticity as the ability of reorganizing metabolic fluxes to maintain viability in response to reaction failures when the growth condition remains guaranteed. On the other hand, functional redundancy applies to the simultaneous use of alternative fluxes in a given medium, even if some can completely or partially compensate for the other~\cite{Wagner:2005}. We perform an exhaustive computational screening of synthetic lethal reaction pairs in {\it E. coli} in a glucose minimal medium and we find that SL reaction pairs divide in two different groups depending on whether the SL interaction works as a backup or as a parallel use mechanism, the first corresponding to essential plasticity and the second to essential redundancy. When comparing the metabolisms of {\it E. coli} and {\it Mycoplasma pneumoniae}, we find that the two organisms exhibit a large difference in the relative importance of plasticity and redundancy. In {\it E. coli}, the analysis of how pathways are entangled through SL pairs supports the view that redundancy SL pairs preferentially affect a single function or pathway~\cite{Hartman:2001}. In contrast --and in agreement with reported SL genetic interactions in yeast \cite{Kelley:2005}-- essential plasticity, which is the dominant class in {\it E. coli}, tends to be inter-pathway but concentrated and unveils Cell Envelope Biosynthesis as an essential backup for Membrane Lipid Metabolism. Finally, different environmental conditions are tested to explore the interplay between these two mechanisms in coessential reaction pairs.

\section{Results}
We use FBA~\cite{Orth:2010} (see Materials and Methods) under glucose minimal medium to compute all SL reaction pairs in the {\it i}JO1366 {\it E. coli} metabolic network reconstruction~\cite{Palsson:2011}, and also in the {\it i}JW145 {\it M. pneumoniae} model~\cite{Wodke:2013,Yus:2009} for a comparative analysis. A reaction pair deletion is annotated as inviable, and so as a synthetic lethal, if the double mutant shows a no-growth phenotype. We preliminarily reduce the space of reactions to be considered in forming potential SL pairs to the set of reactions that can be active but not essential in glucose minimal medium, irrespective of the level of attainable growth (see Materials and Methods). This ensemble, formed of $1176$ reactions in {\it E. coli} (see Supplementary Data Table S1) and $66$ in {\it M. pneumoniae} (see Supplementary Data Table S9), is a subset of the original reconstruction that includes but that is not limited to the set of FBA active reactions under maximum growth constraint~\cite{Schilling:2003,Thiele:2010}. Next, we analyze in detail the classification of identified SL reaction pairs into plasticity and redundancy subtypes (Fig.~\ref{fig:1}).

\subsection{Plasticity and redundancy subtypes of synthetic lethal reactions pairs}

\begin{figure}[t!]
\begin{center}
\includegraphics[width=\textwidth]{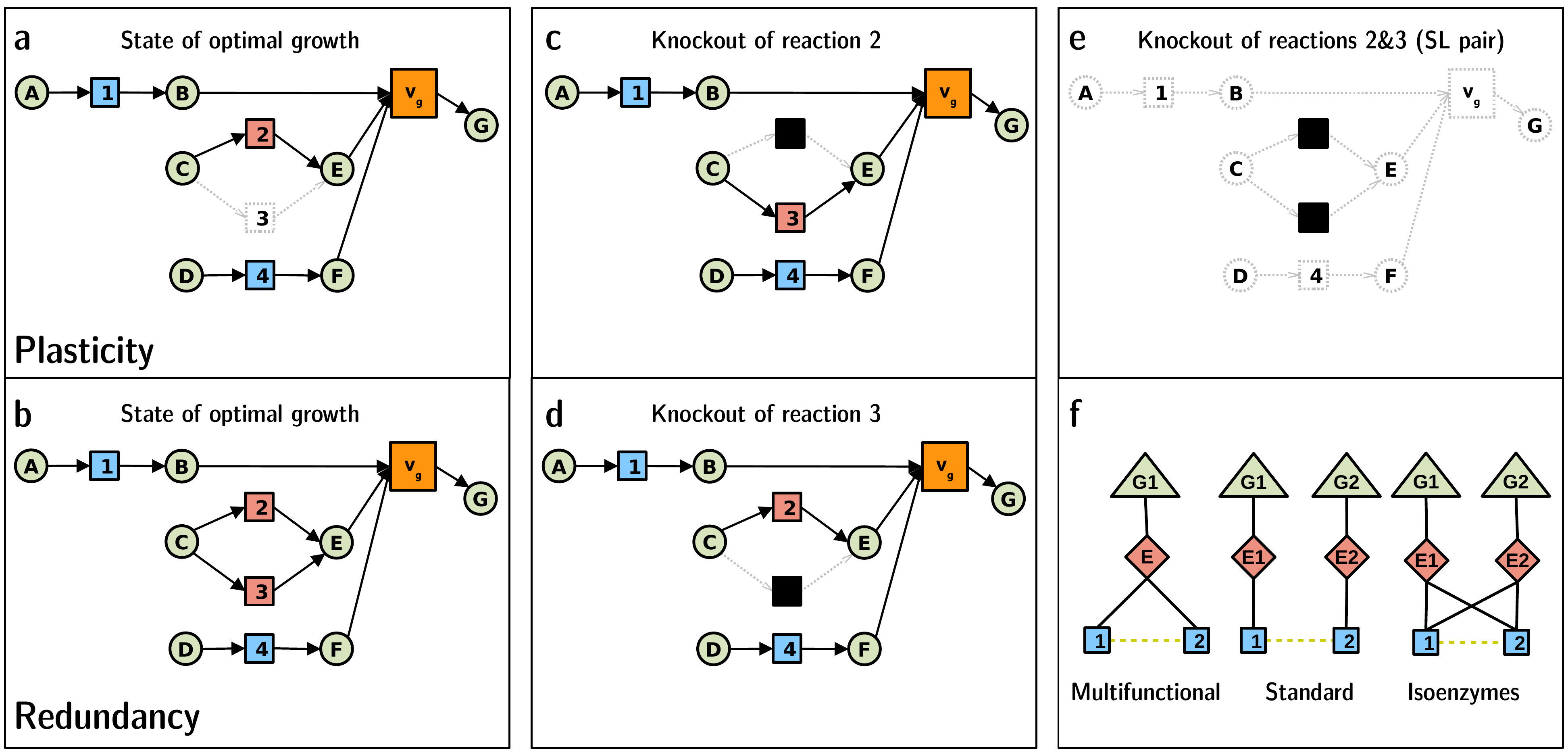} 
\end{center}
\caption{Schematic representation of plasticity and redundancy synthetic lethality subtypes in metabolic networks. Metabolites are represented by circles and reactions by squares. Colored reactions with black arrows represent active reactions, whereas grey discontinuous lines are used for inactive reactions and metabolites and black for knockouts. The biomass production reaction is represented as a larger square with an associated flux $\nu_g$. When it turns to inactive, meaning that it has no associated flux, the organism is not able to grow. For simplicity, SL reaction pairs are illustrated in this figure as having a common metabolite, although this is not necessarily always the case (see text). {\bf a}. Initial configuration of a plasticity synthetic lethality reaction pair (reaction 2 active and reaction 3 inactive).  {\bf b}. Initial configuration of a redundancy synthetic lethality reaction pair (both reactions 2 and 3 active).  {\bf c}. Final configuration after knockout of reaction 2 in {\bf a} or {\bf b}. {\bf d}. Final configuration after knockout of reaction 3 in {\bf a} or {\bf b}.  {\bf e}. Final configuration after simultaneous knockout of reactions 2 and 3 in {\bf a} or {\bf b}. {\bf f}. Different possible organization of genes, enzymes and reactions in SL pairs.}
\label{fig:1}
\end{figure}

We found that $0.04\%$ of all reaction pair deletions in {\it E. coli} are {\it in silico} synthetic lethals and can be separated in two different subtypes. In the biggest group, having a relative size of $91\%$, one of the paired reactions is active in the medium under evaluation while the second reaction has no associated flux. The rest of SL reaction pairs are formed by two active reactions.

As discussed previously in~\cite{Suthers:2009}, some FBA computationally predicted SL pairs can be inconsistent since they contain at least one gene reported as essential {\it in vivo}. In our screening, and in accordance with results in~\cite{Suthers:2009}, this situation corresponds to $4\%$ of all identified SL pairs (see Materials and Methods). Apart from these inconsistencies, active-inactive coessential reaction pairs are referred to as plasticity synthetic lethality (PSL) pairs (Fig.~\ref{fig:1}a). We found $219$ PSL reaction pairs, $86\%$ of all diagnosed SL pairs in {\it i}JO1366 (Fig.~\ref{fig:2}) (see Supplementary Data Table S2). Coessential inactive and active reactions in these pairs have zero and non-zero FBA flux respectively. When the active reaction is removed from the metabolic network, fluxes reorganize such that the zero-flux reaction in the pair turns on as a backup of the removed reaction to ensure viability of the organism, even though the growth is generally lowered. In contrast, the level of growth is unperturbed when the inactive reaction is removed. As an example, the SL pair valine-pyruvate aminotransferase and valine transaminase form a PSL pair, the second reaction being the backup of the first (their simultaneous knockout produces auxotrophic mutants requiring isoleucine to grow~\cite{Whalen:1982}).

\begin{figure}
\begin{center}
\includegraphics[width=0.48\textwidth]{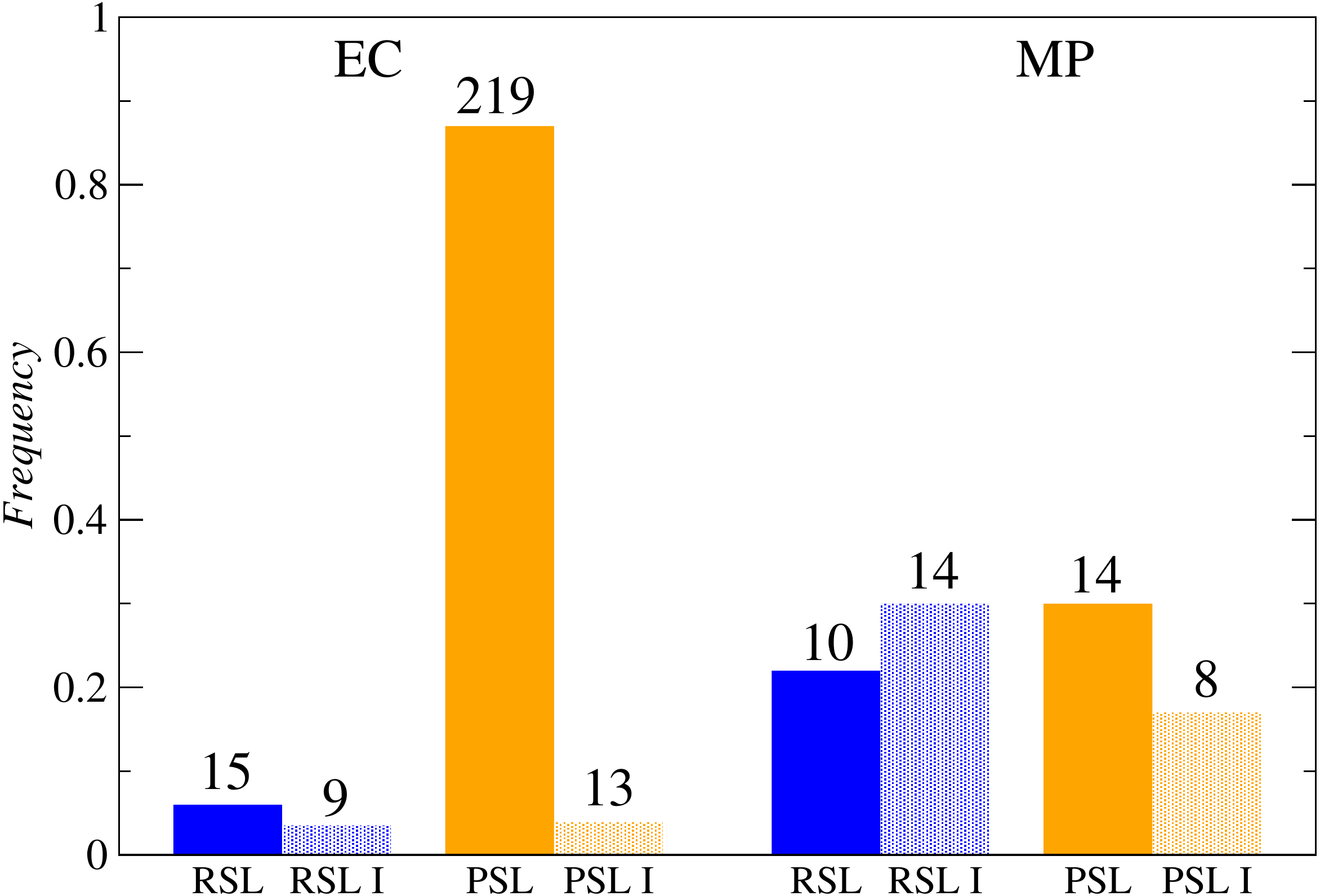}
\end{center}
\caption{Histogram for the four different categories of SL pairs in {\it E. coli} (left) and {\it M. pneumoniae} (right). RSL (blue): redundancy synthetic lethality pairs, RSL I (blue points): redundancy synthetic lethality pairs showing inconsistencies (see text), PSL (orange): plasticity synthetic lethality pairs, PSL I (orange points): plasticity synthetic lethality pairs showing inconsistencies (see text).}
\label{fig:2}
\end{figure}

While the single activation of one of the reactions in a PSL pair is enough to ensure viability in front of single reaction disruptions, the parallel use of both coessential reactions may happen in other cases. We name redundancy synthetic lethality (RSL) reaction pairs those in which both reactions are active and used in parallel (Fig.~\ref{fig:1}b). Of all SL reaction pairs in {\it i}JO1366, we found that $15$ ($6\%$) are RSL (Fig.~\ref{fig:2}) (see Supplementary Data Table S2). We checked indeed that for $13$ of the $15$ RSL pairs the simultaneous use of both reactions increases fitness as compared to the situation when only one of the reactions is active (fitness is here understood as the maximal FBA biomass production rate for the organism). For the remaining two pairs growth remains unchanged. As an illustrative example of parallel use, oxygen transport combines with reactions in the ATP forming phase of Glycolysis to form RSL reaction pairs. If Oxydative Phosphorylation is blocked by the absence of oxygen and no alternative anaerobic process like Glycolysis is used, the energy metabolism of {\it E. coli} collapses and so the whole organism.

Network distance (see Materials and Methods) between reaction counterparts is slightly shorter in RSL pairs than in PSL pairs. Indeed, not all reactions in RSL or PSL pairs are directly connected through common metabolites. Direct connections happen for $60\%$ and $38\%$ of pairs respectively, while the rest can be separated by up to four other intermediate reactions so that the average shortest paths are $3.33$ and $3.80$, respectively (the average shortest path of the whole metabolic network is $5.02$). Both essential plasticity and redundancy display overlap in reactions and associated genes. In the $15$ RSL pairs, we identified $17$ different reactions controlled by $15$ genes or gene complexes. The $219$ PSL pairs involve $108$ different reactions controlled by $61$ genes or gene complexes.

Although our analysis refers to reactions, specific signatures of enzyme activity may be worth stressing in connection with our analysis of coessential reaction pairs. For some of the identified SL pairs, we found direct experimental evidence reported in the literature~\cite{Smith:1993,Whalen:1982} and other results that support the buffering activity of reactions in some SL pairs, like in the aerobic/anaerobic synthesis of Heme \cite{Troup:1995,Rompf:1998} and in the oxidative/non-oxidative working phases of the Pentose Phosphate Pathway \cite{Jiao:2003}. In other cases, we found that enzymatic degeneracy can be responsible for explaining two of the {\it in silico} detected RSL reaction pairs in {\it E. coli}. One RSL reaction pair --that produces isopentenyl diphosphate and its isomer dimethylallyl diphosphate, biosynthetic precursors of terpenes in {\it E. coli} that have the potential to serve as a basis for advanced biofuels \cite{Rude:2009}-- is catalyzed by a single enzyme encoded by an essential gene (one-to-many enzyme multifunctionality (Fig.~\ref{fig:1}f)). Conversely, isoenzymes are encoded by different genes but can catalyze the same biochemical reactions. This many-to-one relationship ensures that single deletion mutants lacking any of the genes encoding one of the isoenzymes can still be viable (Fig.~\ref{fig:1}f). We found that this case happens in $1$ RSL reaction pair catalyzed by isoenzymes encoded by nonessential genes associated to transketolase activity in the Pentose Phosphate Pathway~\cite{Zhao:1994}.

Finally, a comparative study shows that coessential reaction pairs are $50$ times more abundant in a much simpler genome-reduced organisms of increased linearity and reduced complexity such as {\it M. pneumoniae}~\cite{Serrano:2012,Guell:2012}. We found that $2\%$ of all potential candidate reaction pairs in {\it M. pneumoniae} are synthetic lethals (see Supplementary Data Table S10), vs solely the $0.04\%$ in {\it E. coli}. Inconsistencies are also much more abundant relatively to {\it E. coli} and the balance of RSL vs PSL reaction pairs is also different (Fig.~\ref{fig:2}). Parallel use happens as frequently as the backup mechanism in coessential reactions, with $42\%$ of all synthetic lethals being RSL pairs and $58\%$ being PSL pairs. As compared to results reported in~\cite{Wodke:2013} for the synthetic lethality of genes, our methodology detects the same $29$ SL gene pairs and $15$ new SL gene pairs. Since the $8$ different genes in these pairs form two different complexes of $4$ and $3$ genes and one gene remains isolated, the $15$ SL gene pairs reduce to just $2$ SL reaction pairs (in the RSL and RSL I categories) sharing one of the reactions. The $3$ reactions involved in the pairs are uptake of G3P, G3P oxidation to dihydroxyacetone phosphate, and uptake of orthophosphate. As reported in Ref. ~\cite{Wodke:2013}, two independent routes through third-party pathways connect glycolysis to lipid biosynthesis. The first two reactions above, R1 and R2, are involved in one of the routes, while the last reaction R3 influences the flux through the other route. When R1 and R3 or R2 and R3 are removed from {\it i}JW145 model, the organism collapses due to the simultaneous failure of both routes.

\subsection{Pathways entanglement through essential plasticity and redundancy}
\begin{figure}
\begin{center}
\includegraphics[width=\textwidth]{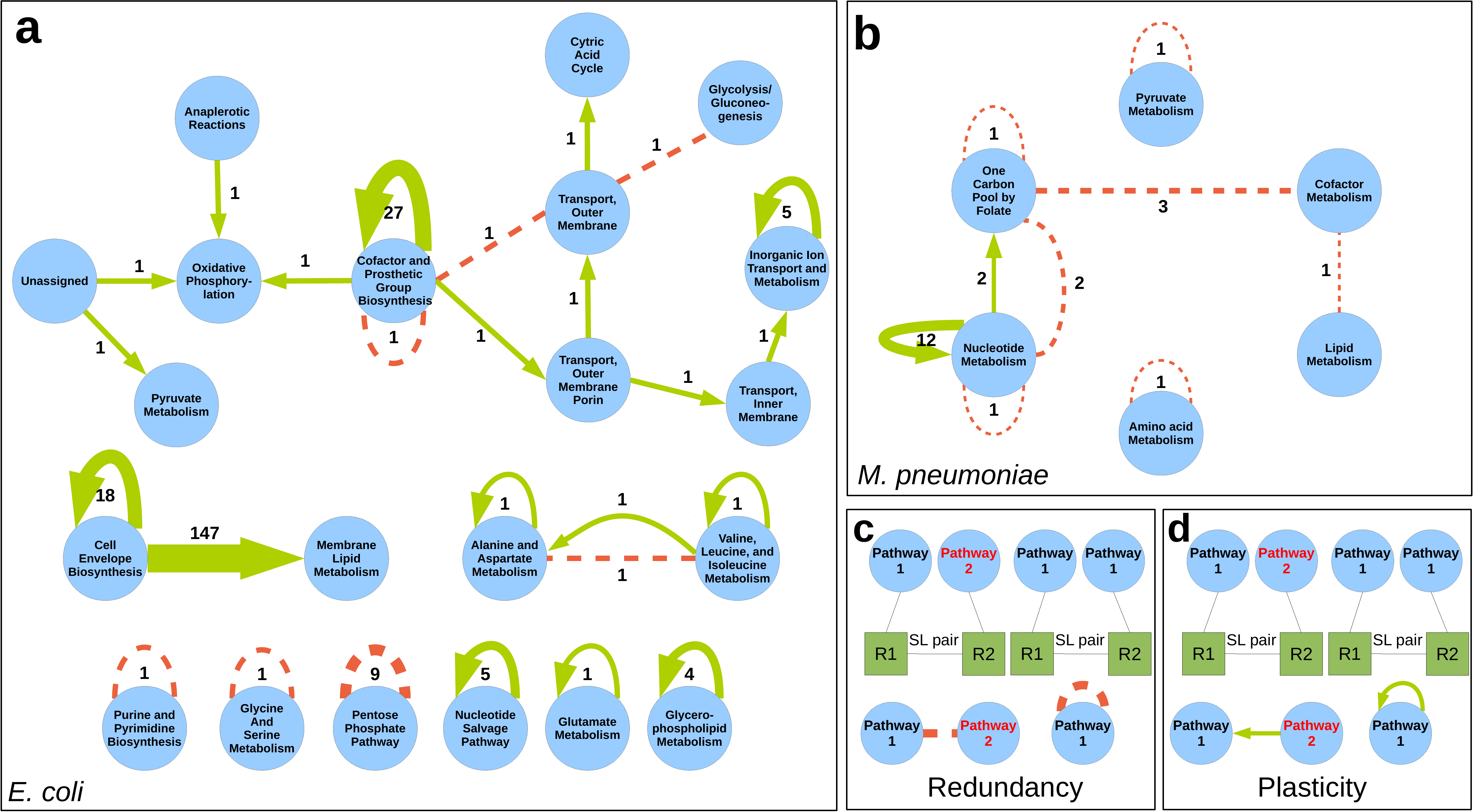}
\end{center}
\caption{Metabolic pathways entanglement through essential plasticity and redundancy in {\it E. coli} and {\it M. pneumoniae}. Nodes represent pathways and two pathways are joined by a link whenever there exists a SL pair containing one reaction in each pathway. Links corresponding to plasticity SL pairs are represented by green continuous arrows pointing from backup to active. Redundancy SL pairs are represented by discontinuous red lines. Labels correspond to the number of pairs which generate this combination of pathways, being thicker those links with more associated pairs. Self-loops correspond to SL pairs with both reactions in the same associated pathway. {\bf a}. Pathways entanglement in {\it E. coli}. {\bf b}. the same for {\it M. pneumoniae}. {\bf c}. Scheme of how pathways entanglement is derived from RSL pairs. {\bf d}. the same for PSL pairs.}
\label{fig:3}
\end{figure}

To investigate further the role of essential plasticity and redundancy in the global organization of metabolic networks, we studied the entanglement of biochemical pathways through synthetic lethality. We annotated all reactions in synthetic lethal pairs in terms of the standard metabolic pathway classification and counted the frequencies of dual pathways combinations both for plasticity and redundancy subtypes. In Fig.~\ref{fig:3}, we give a visual summary of pathways entanglement through essential plasticity and redundancy using a graph representation where pathways are linked whenever they participate together in a SL interaction (discontinuous lines represent redundancy SL interactions (Fig.~\ref{fig:3}c) and continuous arrows stand for plasticity SL interactions (Fig.~\ref{fig:3}d)). The frequency of a given pathway combination in RSL or PSL pairs defines the weight of the corresponding link.

In {\it E. coli} (Fig.~\ref{fig:3}a), we observe that the synthetic lethality entanglement of pathways is in general very low, with the exception of the entanglement between Cell Envelope Biosysthesis and Membrane Lipid Metabolism. Redundancy SL pairs are basically intra-pathway, with only 3 of 15 being inter-pathway. Of all intra-pathway RSL pairs, $75\%$ concentrate in the Pentose Phosphate pathway. Interestingly, the distribution of PSL reaction pairs avoids that of RSL pairs and, in contrast, tends to be inter-pathway. Of all PSL pairs, $67\%$ include zero-flux reactions in Cell Envelope Biosysthesis and active reactions in the Membrane Lipid Metabolism, which unveils Cell Envelope Biosysthesis as an essential backup for Membrane Lipid Metabolism. Intra-pathway plasticity coessentiality amounts to $29\%$ of PSL pairs and is concentrated in Cofactor and Prosthetic Group and Cell Envelope Biosynthesis. 

In {\it M. pneumoniae} (Fig.~\ref{fig:3} b) pathways entanglement through coessentiality of reactions is very low as in {\it E. coli}. Redundancy SL pairs can be intra-pathway ($4$ of $10$) or inter-pathway ($6$ of $10$) and PSL pairs are basically intra-pathway ($12$ of $14$). Redundancy SL pairs denote the parallel use of reactions in Folate Metabolism and reactions in Nucleotide and Cofactor Metabolism. These two pathways, Folate and Nucleotide Metabolism, are also linked by $2$ PSL pairs with non-essential reactions in Folate Metabolism and essential reaction backups in Nucleotide Metabolism. Nucleotide Metabolism is also the pathway that concentrates most PSL pairs. Both RSL and PSL reaction pairs unveil Nucleotide and Folate Metabolism as the most entangled pathways.

\subsection{Sensitivity to environmental conditions}
We finally investigate whether the subtype of a coessential pair switches between plasticity and redundancy depending on the growth condition under evaluation. Environmental specificity of genes and reactions has been explored experimentally~\cite{Palsson:2011,Giaever:2002,Steinmetz:2002} and {\it in silico}~\cite{Barve:2012} for different organisms and for random viable metabolic network samples, and it has also been extended to multiple knockouts in yeast \cite{Deutscher:2006,Harrison:2007} and {\it E. coli}~\cite{Nakahigashi:2009}.

We investigate the sensitivity of SL reaction pairs in {\it E. coli} to changes in minimal medium composition. We focus on the $234$ SL pairs detected in glucose minimal medium, and we check their classification over $333$ minimal media (formed of mineral salts and one variable source of carbon, nitrogen, sulfur and phosphorus~\cite{Palsson:2011}, see Materials and Methods). In Fig. 4a, we display the SL reaction pairs ranked by the fraction of media in which the pairs are synthetically lethal. We find that for most pairs, coessentiality is not specific of an environment and only a minimal number of pairs shows environmental specificity. In particular, $53\%$ coessential pairs are lethal in all media and $95\%$ are lethal in more than $95\%$ of environments. For each SL pair, we count the number of media in which the SL pair is classified in the plasticity subtype as compared to the total number of media in which the pair is predicted to be coessential. Results are shown in Fig. 4b. Nearly all SL pairs, $93\%$, are in the plasticity subclass for more than $93\%$ of the media, while $12$ pairs display a switching behavior between plasticity and redundancy. Noticeably, these pairs are intra-pathway and share common metabolites. Of them, $3$ pairs contribute to biosynthesis of amino acids (Valine, Leucine, and Isoleucine Metabolism and Glycine and Serine Metabolism) and $5$ pairs belong to the Pentose Phosphate Pathway and are related to the production of carbon backbones used in the synthesis of aromatic amino acids. Finally, $5$ reaction pairs maintain the redundancy subclass across all conditions in which are coessential. 

\begin{figure}[!ht]
\begin{center}
\includegraphics[width=0.48\textwidth]{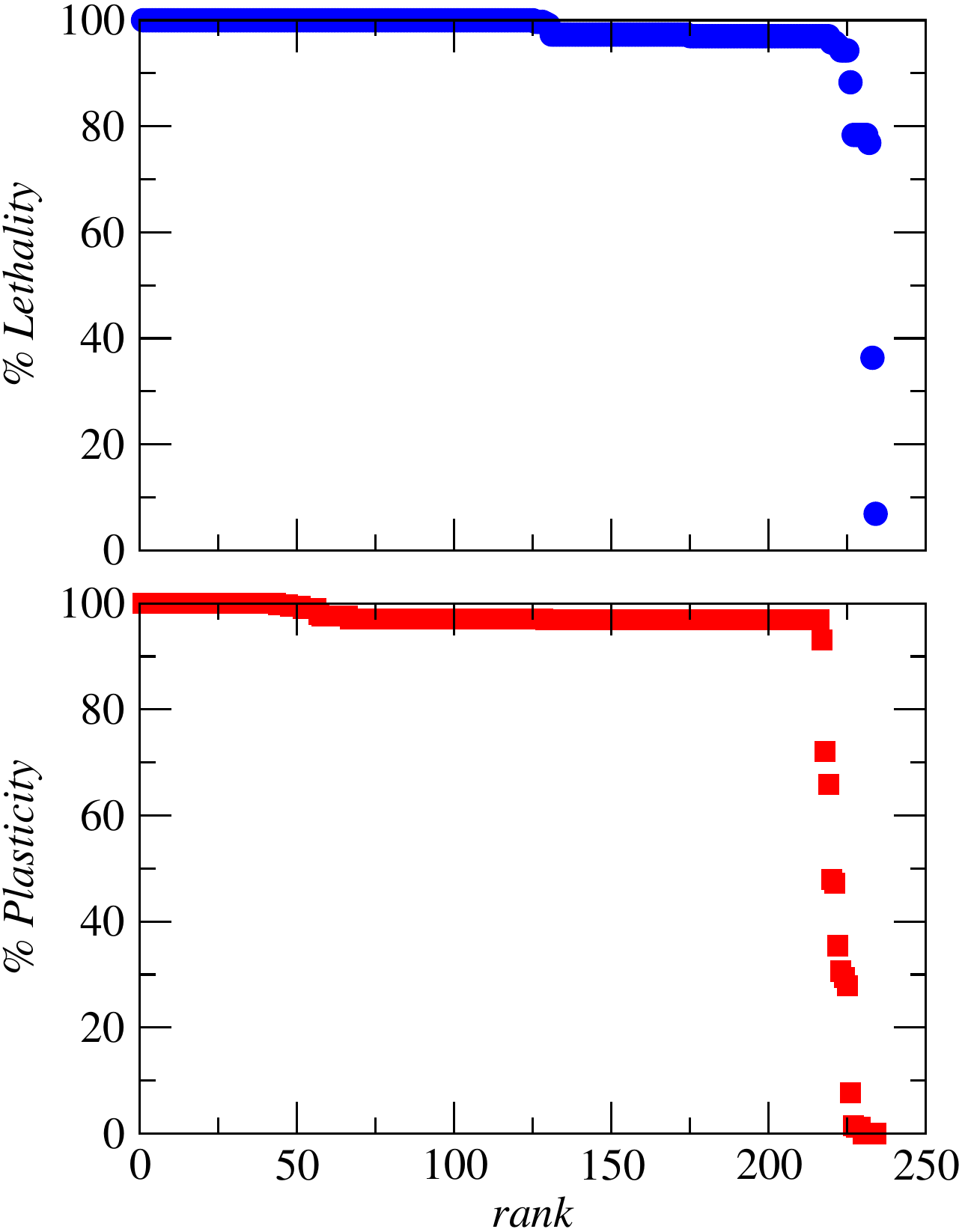}
\end{center}
\caption{Synthetic lethal reaction pairs in minimal media. Top: Synthetic lethal reaction pairs ranked by the fraction of minimal media for which the pair is synthetically lethal. Bottom: Synthetic lethal reaction pairs ranked by the fraction of minimal media in which the SL pairs are classified as essential plasticity and, complementary, as essential redundancy, provided that the pairs remain synthetically lethal.}
\label{fig:4}
\end{figure}

We also explored the behavior of {\it E. coli} in amino acid-enriched medium (see Materials and Methods). Comparing with glucose minimal medium, our first observation is that 223 of the 234 SL pairs detected in glucose minimal medium are also found to be lethal in amino acid-enriched medium (see Supplementary Data Table S3), which means that 11 pairs are rescued. Of the 11 RSL pairs in amino acid-enriched medium, 8 are conserved and 3 switch from plasticity in the minimal to redundancy in the amino acid-enriched medium. On the other hand, $208$ of the $212$ PSL pairs are conserved and 4 change from redundancy in the minimal to plasticity in the amino acid-enriched medium. Noticeably, only in $1$ of the $208$ conserved PSL pairs the pattern of activity changes from the reductase reaction producing dimethylallyl diphosphate to the isomerization of the less reactive isopentenyl pyrophosphate. In addition, a set of lethal reaction pairs ($12195$) occur, all of them involving however one essential reaction in glucose minimal medium that in amino acid-enriched medium becomes nonessential and instead takes part in a SL pair. Apart from those, no other new SL pairs are found. In addition, we redid our simulations taking into consideration a LB based rich medium~\cite{Wunderlich:2006,Sezonov:2007} (see Supplementary Data Table S4). In this rich medium, we only found $13$ new rescues when compared to the minimal medium (two new rescues as compared to the amino acid-enriched medium) and only $3$ SL pairs change their plasticity/redundancy category (see Supplementary Data Table S5).

We found that plasticity and redundancy are still conserved when the growth maximization requirement is loosen (see Materials and Methods). If growth is relaxed in {\it E. coli} to $30\%$ of its maximum value in glucose minimal medium, {\it in silico} essentiality of individual reactions does not change but activation of reactions increases. We find however that the effect of this reorganization is indeed mild for plasticity and redundancy. When biomass production is relaxed by overconstraining the upper bounds of the uptake rates of mineral salts, all SL pairs are conserved and $82\%$ of them maintain their PSL or RSL classification. The absolute number of RSL pairs increases from $15$ to $50$ since $4$ of RSL pairs in the reference conditions given by glucose minimal medium change to plasticity in the overconstrained medium, and at the same time $39$ PSL pairs change to RSL. On the other hand, $180$ SL pairs of $219$ in the reference medium remain as PSL pairs in the overconstrained condition. However, the pattern of activity in the pair has switched in $14\%$ of the PSL pairs in this case, which indicates that the specific selection of the active reaction in a PSL pair can have an impact in the level of attainable growth (see Supplementary Data Table S6). If instead of limiting the uptake of mineral salts we overconstrain the uptake rates of basic nutrients providing sources of carbon, nitrogen, phosphorus and sulfur, the effect is even softer and indeed negligible as compared to the reference medium. The number of active reactions only increases in $3$, the essentiality of individual reactions and SL pairs is conserved, and $99\%$ of them maintain their PSL or RSL classification with only $3$ SL pairs that switch class and only $1$ PSL pair that changes the active reaction (see Supplementary Data Table S7).

\section{Discussion}
Synthetic lethals are complex functional combinations of genes or reactions that denote at the same time both vulnerability in front of double deletions and robustness in front of the failures of any of the two counterparts. Although synthetic lethal genes could be associated to the plasticity and redundancy categories, approaching directly pairs of reactions without the (sometimes multifunctional) scaffold of enzymes and genes allows us to determine in a clean and systematic way the minimal combinations of reactions that turn out to be essential for an organism. Working at the level of reactions, we showed that this synthetic lethality is meditated by two different mechanisms, essential plasticity and essential redundancy, depending on whether one reaction is active for maximum growth in the medium under consideration and the second inactive, or in contrast both reactions have non-zero flux. Plasticity sets up as a sophisticated backup mechanism (mainly inter-pathway in {\it E. coli}) that is able to reorganize metabolic fluxes turning on inactive reactions when coessential counterparts are removed in order to maintain viability in a specific medium, while redundancy corresponds to a simultaneous use of different flux channels (mainly intra-pathway in {\it E. coli}) that ensures viability and besides increases fitness. Apparently, it could seem extremely improbable that the removal of an inactive reaction together with a non-essential active one, like in PSL pairs, could have any lethal effect on an organism. However, we found that this situation is indeed overwhelmingly dominant in {\it E. coli} as compared to redundancy synthetic lethality, and it is still relatively frequent even in a less complex organism like {\it M. pneumoniae}. 

Synthetic lethal mutations have been assumed to affect a single function or pathway~\cite{Hartman:2001}, which reinforces the idea that pathways act as autonomous self-contained functional subsystems. In contrast, other investigations in yeast \cite{Kelley:2005} report that synthetic-lethal genetic interactions are approximately three and a half times as likely to span pairs of pathways than to occur within pathways. In our work, we found that RSL pairs in {\it E. coli} are predominantly intra-pathway while PSL pairs, more abundant, tend to be inter-pathway although concentrated in the entanglement of just two pathways, Cell Envelope Biosynthesis and Membrane Lipid Metabolism. The comparative study here shows that although pathways entanglement through coessentiality of reactions is low in both organisms, RSL pairs in {\it M. pneumoniae} can be intra-pathway or inter-pathway, linking Folate Metabolism and Nucleotide and Cofactor Metabolism, and PSL pairs are basically intra-pathway and located in Nucleotide Metabolism. Taken together, these results indicate that Folate and Nucleotide Metabolic pathways preserve most rescue routes for reaction deletion events, in accordance with results in~\cite{Wodke:2013}. The fact that the proportion of plasticity SL pairs is considerably decreased in {\it M. pneumoniae} as compared to {\it E. coli} could be indicative that, even if both plasticity and redundancy serve an important function in achieving viability, essential plasticity is a more sophisticated mechanism that requires a higher degree of functional organization, using at the same time less resources for maximum growth. At the same time, this can also be explained by the relative unchanging environmental conditions of {\it M. pneumoniae} in the lung, that could have induced the elimination of pathways not required in that medium~\cite{Wodke:2013}. This suggests that the adaptability of {\it M. pneumoniae} is very much reduced and its SL behavior could not be resilient to environmental changes.

We also found that SL reaction pairs and their subdivision in plasticity and redundancy are highly conserved independently of the composition of the minimal medium that acts as environmental condition for growth, and even when this environment is enriched with nonessential compounds or overconstrained to decrease the maximum biomass production. These environment unespecific SL pairs can thus be selected as potential drug targets operative regardless of the chemical environment of the cell. We admit that large-scale computational screenings might be biased both by model details and by the quality of the experimental data used to feed the metabolic reconstruction, which could affect the identification of SL pairs and their categorization in plasticity and redundancy. However, we believe that FBA and the genome-scale metabolic reconstructions that we use in this work have proven to be reliable and highly congruent with empirical observations and that, with few exceptions, our results will pass the test of new modeling approaches that may become the reference in the future.

Our work was intended as an advancement in setting the basis for an analysis of essential plasticity and essential redundancy in metabolic networks. The elucidation of these capacities through synthetic lethal interactions has clear implications for biotechnology and biomedicine, since targeting a PSL or a RSL pair will certainly have different requirements and implications. In human cells, the analysis of PSL and RSL pairs can help the design of new drug targets, once we solve the enormous tasks challenging the accurate understanding of cell metabolism in humans, {\em e.g.}, the lack of association between the majority of genes and a recognized function and the deficiency in reconstruction models for each specific human cell type. Beyond metabolic networks, plasticity and redundancy are also very important concepts for other biological complex systems, like the brain. Evolution has been postulated as the main explanation for the plasticity and redundancy abilities observed in the brain, with the hypothesis that when new parts of the brain evolved and took over the old brain functions, the old areas that mediated those functions still retained some control \cite{Somel:2013}. Whether essential plasticity and essential redundancy are adaptive in cell metabolism or, as it has been argued for metabolism in changing environments~\cite{Harrison:2007, Wang:2009}, they are rather a byproduct of the evolution of biological networks toward survival, these regulatory mechanisms are key to understand how complex biological systems protect themselves against malfunction.

\section{Materials and Methods}
\subsection{Identification of the space of potential reactions in SL pairs: biomass unconstrained Flux Variability Analysis}
Preliminary to the identification of SL pairs using FBA, we filter the set of reactions in the metabolic reconstruction of the organisms to keep only those potentially relevant. As a first step, we only take into account reaction pairs with associated genetic information. Hence, spontaneous reactions or reactions with unknown associated genes are discarded, leading to the elimination of 117 reactions for {\it E. coli} and 48 reactions for {\it M. pneumoniae}.

To identify potential synthetic lethal reaction pairs, first we compute which reactions can have a non-zero flux in a particular medium composition (glucose minimal medium/amino acid-enriched medium/rich medium). To do this, we follow an approach similar to Flux Variability Analysis (FVA) \cite{Thiele:2010}. FVA consists on computing the minimum and maximum values of the fluxes that each reaction can have, taking into account that the growth rate must be fixed to a value that ensures viability. However, being conservative, we are interested in capturing all the possible scenarios independently of the value of the flux of the biomass reaction, since in this way we can take also into account non-optimal/low-growth scenarios. Therefore, we modify FVA to compute the minimum and maximum possible values of the flux of each reaction regardless of the value of the biomass formation rate. To this end, we do not constrain the value of the flux of the biomass reaction and we just allow any positive value, $\nu_{biomass} \geq 0$. Under this condition, we determine which reactions have maximum or minimum values of the fluxes different from $0$. By doing this we obtain the maximal set of reactions which can be active in the particular medium under consideration independently of the rate of biomass formation of the organism.

\subsection{Identification of synthetic lethal reaction pairs: Flux Balance Analysis}
Flux Balance Analysis is a method to compute the fluxes of reactions in metabolic networks at steady state. In its standard version, FBA optimizes biomass production without using kinetic parameters~\cite{Orth:2010}. This method is able to predict the growth rate and the fluxes of a metabolic network~\cite{Schilling:2000,Schilling:1998} with high accuracy. In particular, FBA was shown to predict gene essentiality with an accuracy of 90\% \cite{Feist:2007, Wang:2009}. It was also used for {\it in silico} prediction of SL pairs~\cite{Suthers:2009,Deutscher:2006}.

We implement FBA using GNU Linear Programming Kit (GLPK) to compute synthetic lethality of reaction pairs. Once we have determined the space of reactions to be considered in potential SL pairs using FVA (see Materials and Methods subsection above ``Identification of the space of potential reactions in SL pairs: biomass unconstrained Flux Variability Analysis''), we apply an exhaustive search checking all potential pairs. For each possible combination, we compute FBA on the mutant obtained by removing the corresponding pair of reactions from the metabolic network model. We annotate the double deletion as inviable, and so as a synthetic lethality pair, if FBA shows a no-growth phenotype. Notice that reversible reactions are treated as two coupled reactions to account for the forward and reverse fluxes.

In order to validate our methodology (including the FVA reaction set selection), we computed {\it in silico} single essential reactions and SL reaction pairs for the iAF1260 model of {\it E. coli} in glucose minimal medium. Our results are in perfect agreement with those reported in~\cite{Suthers:2009}.

\subsection{Identification of inconsistencies}
Pairs, formed by reactions, both having an associated gene or genetic entity, are checked according to their experimental essentiality. For {\it E. coli}, we use the information given in Ref. \cite{Palsson:2011}. For {\it M. pneumoniae}, we use results from a genome-wide transposon study in {\it M. genitalium} given in Ref.~\cite{Glass:2006}. A functional ortholog in {\it M. genitalium} can be assigned to 128 metabolic genes in {\it i}JW145, and we associate the essentiality of that ortholog to the corresponding gene in {\it M. pneunomiae}. The other $17$ genes are assumed (similarly to Ref.~\cite{Wodke:2013}) to be not essential for growth due to their absence in {\it M. genitalium} and the high similarity of the metabolic networks of both organisms (Ref.~\cite{Yus:2009}). Three cases may occur:

\begin{itemize}
 \item Both reactions have non-essential genes. In this case, the pair can be considered RSL or PSL.
 \item One reaction has a non-essential gene whereas the other has an essential one. In this case, we focus on the essential one. If the latter regulates more than one reaction, we consider that the pair is not an inconsistency, since the essentiality might refer to the rest of regulated reactions. Otherwise, the pair is considered as inconsistent.
 \item Both reactions are regulated by essential genes. With the same argument as before, for the case that both reactions have associated genes which regulate more than one reaction, we consider the pair able to be RSL or PSL. The other possible combinations are considered inconsistencies.
\end{itemize}
Besides, detected SL pairs with an active and an inactive reaction (PSL pairs) associated to isoenzymes (2 occurrences) and multifunctional enzymes (1 occurrence) are classified as inconsistencies.

\subsection{Computation of shortest path lengths}
We model metabolism as a bipartite directed network \cite{Guell:2012}, where directed links connect metabolites with reactions in which they participate as reactants or products. The shortest path length measures the topological distance between reactions in this network representation. It is calculated as the minimal number of different intermediate nodes (metabolites and reactions) visited when going from one reaction to another following the directed links. In practice, we use Dijkstra's Algorithm \cite{Dijkstra:1959}. to compute these network-based distances.

\subsection{Simulation of environmental conditions}
\begin{itemize}
\item {\bf {\em E. coli} minimal media.} We considered the set of 333 minimal media given in Ref. \cite{Palsson:2011}. To construct them, a set of mineral salts (always the same) and 4 extra metabolites, each one representing a source of carbon, nitrogen, phosphorus and sulfur respectively, is included. To set the sources of basic elements, first a particular element is chosen, for instance carbon, and the metabolite containing that source was varied within a group (in the case of carbon, candidates are glucose, fructose, etc.) while keeping the metabolites which are the sources of the remaining elements fixed to the reference choice (C: glucose, N: ammonium, P: phosphate, S: sulfate). The procedure is repeated for the remaining basic elements. The number of media constructed in this way was 555. However, a final number of 333 viable media was retained after computing FBA.

\item {\bf {\em E. coli} amino acid-enriched medium.} Amino acid-enriched medium exchange bounds have not been measured yet for the version of {\it E. coli} that we use in this work. However, an amino acid-enriched medium \cite{Orth:2013} can be constructed from the glucose minimal medium given in \cite{Palsson:2011} by adding the following set of amino acids: D-Alanine, L-Alanine, L-Arginine, L-Asparagine, L-Aspartate, D-Cysteine, L-Cysteine, L-Glutamine, L-Glutamate, Glycine, L-Histidine, L-Homoserine, L-Isoleucine, L-Leucine, L-Lysine, L-Methionine, L-Phenylalanine, L-Proline, D-Serine, L-Serine, L-Threonine, L-Tryptophan, L-Tyrosine, L-Valine. This set of amino acids enriches the glucose minimal medium allowing the organism to take them as nutrients. Otherwise the organism would have to synthesize them, resulting in a more stringent environment for the organism. To simulate the presence of this set of amino acids in the medium, we set the exchange constraints bounds of these amino acids to -10 mmol/(gDW$\cdot$h) ($\nu^{exchange}_{amino\: acid} \geq -10$).

\item {\bf {\em E. coli} rich medium.} To construct a rich medium we take into consideration a Luria-Bertani Broth \cite{Sezonov:2007}, which contains as additional compounds purines and pirimidines apart from amino acids. We also added vitamins, namely biotin, pyridoxine, and thiamin, and also the nucleotide nicotinamide monocleotide \cite{Wunderlich:2006}. Other compounds, like PABA or chorismate, cannot be uptaken by the {\it E. coli} model we are using. The exchange constraints bounds of these compounds are set to -10 mmol/(gDW$\cdot$h) ($\nu^{exchange}_{compound} \geq -10$). A detailed list of the added compounds is given in our Supplementary Data Table S4.

\item {\bf Overconstraining nutrients for decreasing biomass production in {\em E. coli}.} To implement the relaxation of the growth maximization requirement we took as a reference the glucose minimal medium given in \cite{Palsson:2011} and limited the biomass production or the basic nutrients uptake rates. In the first case, we performed a FVA calculation fixing the growth of the biomass to $30\%$ of the maximal growth in glucose minimal medium and obtained the exchange bounds of all nutrient uptakes. In comparison to the reference values, we observe that the only metabolites which lower their maximal uptakes are the mineral salts (approximately reduced also to a $30\%$), while the uptake rates for the rest of compounds remained with the same bounds. We then performed FBA calculations in this overconstrained condition and computed SL pairs and their classification in RSL and PSL. 

We also checked the effect of constraining the uptake rates of the four basic compounds containing sources of carbon, nitrogen, phosphorus, and sulfur instead of those of the mineral salts. We first applied FVA setting the value of biomass growth to the maximum in glucose minimal medium in order to determine an upper uptake limit. Then, we constrained the maximum rate uptake of glucose and of the other three basic compounds to $30\%$ of the maximum possible values while keeping the reference values for the mineral salts. We applied FBA in the resulting overconstrained medium and computed SL pairs and their classification in RSL and PSL. 

In both overconstrained modifications of the glucose minimal medium, the number of active reactions changed from $412$ to $490$ in the mineral salts overconstrained medium and to $415$ in the basic nutrients overconstrained medium. In both cases, the essentiality of individual reactions and all SL pairs were conserved (except for $2$ new RSL inconsistencies in the basic nutrients overconstrained medium).

\item \textit{\textbf{M. pneumoniae medium composition}}. We consider the medium given in Table S5 of the Supplementary Information of Ref. \cite{Wodke:2013}. We use the constraints corresponding to the category called defined medium and we choose to add also D-ribose. In our Supplementary Data Table S7, we give a list with the metabolites present in the environment that we consider in our SL computations.
\end{itemize}

\section{Acknowledgments}
\begin{acknowledgments}
We thank Judith Wodke for solving doubts about the network of {\it M. pneumoniae}. This work has been partially supported by MICINN Projects No. FIS2006-03525, FIS2010-21924-C02-01 and BFU2010-21847-C02-02; and Generalitat de Catalunya grant No. 2009SGR1055. O. G. acknowledges support from the Spanish Ministerio de Ciencia e Innovaci\'on through a FPU grant. M. A. S. acknowledges the Ram\'on y Cajal program of the Spanish Ministerio de Ciencia e Innovaci\'on and the James S. McDonnell Foundation.
\end{acknowledgments}

\bibliography{template}

\end{document}